\documentstyle[multicol,prl,aps,epsfig]{revtex}
\begin{document}
\draft
\title{
Experimental simulation of a stellar photon bath by bremsstrahlung:
the astrophysical $\gamma$-process
}
\author{
  P.~Mohr,$^1$ K.~Vogt,$^1$ M.~Babilon,$^1$ J.~Enders,$^1$ 
  T.~Hartmann,$^1$ C.~Hutter,$^1$ T.~Rauscher,$^{2,3}$ S.~Volz,$^1$ 
  A.~Zilges$^1$
}
\address{$^1$
  Institut f\"ur Kernphysik, Technische Universit\"at Darmstadt,
  Schlossgartenstrasse 9, D--64289 Darmstadt, Germany
}
\address{$^2$
  Institut f\"ur Physik, Universit\"at Basel, 
  Klingelbergstrasse 82, CH-4056 Basel, Switzerland
}
\address{$^3$
  Department of Astronomy and Astrophysics, UCSC,
  Santa Cruz, CA 95064, USA
}
\date{\today}
\maketitle
\begin{abstract}
The nucleosynthesis of heavy proton-rich nuclei in a
stellar photon bath at temperatures of the
astrophysical $\gamma$-process
was investigated
where the
photon bath was simulated by the
superposition of brems\-strahlung spectra with
different endpoint energies. 
The method was applied to derive
($\gamma$,n) cross sections and
reaction rates for several platinum isotopes.
\end{abstract}


% 25.20.-x  Photonuclear reactions
% 26.30.+k  Nucleosynthesis in novae, supernovae and other explosive
%           environments
% 98.80.Ft  Origin and formation of the elements
% 26.45.+h  Elemental and isotopic abundances and evolution

\begin{multicols}{2}
\narrowtext

The trans-iron nuclei have been synthesized
by neutron capture in the $s$- and $r$-processes,
except the $p$-nuclei ($p$ for proton-rich), with
relative abundances of the order of 0.01 to 1\% \cite{Lam92}.
The main production mechanism of the $p$-nuclei
is assumed to be 
photodisintegration in the $\gamma$-process,
i.e.\ by
($\gamma$,n), 
($\gamma$,p), and
($\gamma$,$\alpha$) reactions
induced on heavier seed nuclei synthesized in the
$s$- and $r$-processes. 
Typical parameters for the $\gamma$-process are temperatures of
$2 \le T_9 \le 3$ ($T_9$ is the temperature in units of $10^9$\,K), 
densities $\rho \approx 10^6$\,g/cm$^3$, and time scales $\tau$
in the order of seconds. 
Several astrophysical sites 
for the $\gamma$-process have been proposed, 
whereby the oxygen- and neon-rich layers of type II supernovae
seem to be good candidates. However, no
definite conclusions have been reached yet \cite{Lam92},
predominantly due to the lack of experimental data
for the cross sections and reaction rates 
of the $\gamma$-induced reactions
at astrophysically
relevant energies. All reaction rates have been derived theoretically
using statistical model calculations
\cite{Lam92,Arn99,Lan99,Wal97,Woo78,Ray95}.

The energy distribution of a thermal photon bath at a temperature $T$
is given by the Planck distribution
\begin{equation}
n_\gamma(E,T) = 
  \left( \frac{1}{\pi} \right)^2 \,
  \left( \frac{1}{\hbar c} \right)^3 \,
  \frac{E^2}{\exp{(E/kT)} - 1}
\label{eq:planck}
\end{equation}
where $n_\gamma(E,T)$ is the number of $\gamma$-rays 
at energy $E$ per unit of volume and
energy interval. In a photon-induced reaction
B($\gamma$,x)A the distribution 
leads to a temperature dependent decay rate $\lambda(T)$
of the initial nucleus B
\begin{equation}
\lambda(T) =
  \int_0^\infty 
  c \,\, n_\gamma(E,T) \,\, \sigma_{(\gamma,{\rm{x}})}(E) \,\, dE
\label{eq:gamow}
\end{equation}
with the speed of light $c$ and the cross section of the $\gamma$-induced
reaction $\sigma_{(\gamma,{\rm{x}})}(E)$.
Obviously, $\lambda$ is also the production rate of the residual
nucleus A. In the following we will focus on 
photodisintegration by the ($\gamma$,n) reactions.

\begin{figure}
\epsfig{figure=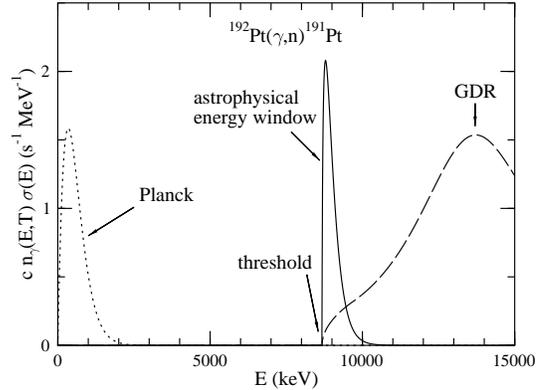,bbllx=100,bblly=85,bburx=380,bbury=285,
width=7.5cm}
%width=15.0cm}
\caption{
  \label{fig:gamow} 
  Astrophysically relevant energy window for 
  the $^{192}$Pt($\gamma$,n)$^{191}$Pt reaction 
  ($E_{\rm{thr}} = 8676$\,keV) in a thermal
  photon bath with temperature $T_9 = 2.5$. 
  The Planck distribution $n_\gamma(E,T)$ (dotted line) and
  the ($\gamma$,n) cross section $\sigma(E)$ (dashed line)
  are given in relative units.
  The product $c \, n_\gamma(E) \cdot \sigma(E)$ shows a maximum at
  the effective energy $E_{\rm{eff}} \approx E_{\rm{thr}} + \frac{1}{2} kT$
  (full line).
  The GDR parameters were taken from experimental data 
  \protect\cite{Die88} and the threshold behavior 
  $\sigma \sim \sqrt{E-E_{\rm{thr}}}$ was matched to the 
  Lorentzian shaped GDR cross section 1\,MeV
  above the threshold. 
}
\end{figure}

A large number of ($\gamma$,n) cross sections has been
measured over the years \cite{Die88,CDFE}. However, most of the data
have been obtained around the giant dipole resonance (GDR),
i.e.\ at energies
much higher than those in stars, and practically no data exist
for the $p$-nuclei. 
The integrand in Eq.~(\ref{eq:gamow})
is given by the product
of the $\gamma$ flux $c \, n_\gamma(E,T)$, 
which decreases steeply
with increasing energy $E$,
and the cross section 
$\sigma_{(\gamma,{\rm{x}})}(E)$, which increases with $E$
approaching the GDR region.
The product leads then to a window 
at an effective energy $E_{\rm{eff}}$ with a width $\Delta$
similar to the Gamow window for charged-particle-induced reactions.
If one assumes a typical threshold behavior of the ($\gamma$,n)
cross section 
close to the threshold energy $E_{\rm{thr}}$,
the effective energy is approximately given by
$E_{\rm{eff}} = E_{\rm{thr}} + \frac{1}{2}kT$,
and the typical width $\Delta$ is in the order of 1\,MeV
(Fig.~\ref{fig:gamow}).

\begin{figure}
\epsfig{figure=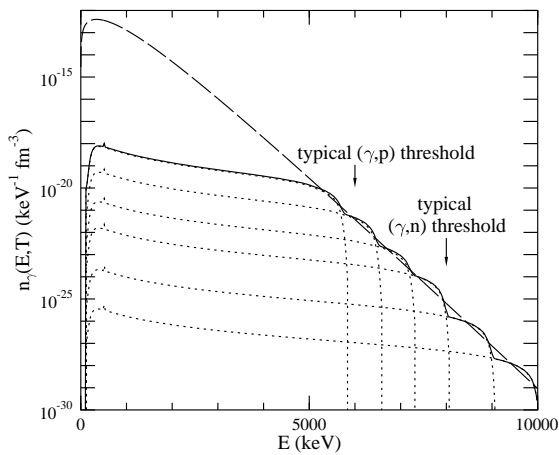,bbllx=100,bblly=85,bburx=380,bbury=315,
width=7.5cm}
%width=15.0cm}
\caption{
  \label{fig:super} 
  The superposition of brems\-strahlung spectra 
  $\Phi_{\rm{brems}}^{\rm{qt}}(T)$ (full line)
  with
  different endpoint energies $E_{0,i}$ is compared with
  the Planck spectrum 
  $n_\gamma(E,T)$ (dashed line)
  at the temperature $T_9 = 2.5$.
  Good agreement is found at $E = $ 6 to 10\,MeV with the superposition
  of six brems\-strahlung 
  spectra 
  $\Phi_{\rm{brems}}(E_{0,i})$
  which are shown as dotted lines.
}
\end{figure}

The energy distribution of brems\-strahlung is approximately described by the 
Schiff formula \cite{Sch51}.
Close to the endpoint energy $E_0$ (which is also the energy of the
incoming electron beam) the brems\-strahlung spectrum decreases steeply
with increasing energy. We found that over a narrow 
energy region the brems\-strahlung
spectrum has a similar shape as the Planck spectrum.
However, a superposition of several brems\-strahlung spectra 
$\Phi_{\rm{brems}}(E_{0,i})$
with different endpoint energies $E_{0,i}$ 
leads to a quasi-thermal spectrum 
$\Phi_{\rm{brems}}^{\rm{qt}}(T)$
which has nearly the same shape as the Planck spectrum 
over a relatively broad energy range (Fig.~\ref{fig:super}):
\begin{equation}
  c \, n_\gamma(E,T) \approx \Phi_{\rm{brems}}^{\rm{qt}}(T)
  = \sum_i a_i(T) \, \cdot \, \Phi_{\rm{brems}}(E_{0,i})
\label{eq:approx}
\end{equation}
where the strength coefficients $a_i(T)$ must be adjusted
for each temperature $T$.
The example shown in Fig.~\ref{fig:super} demonstrates that a reasonable
agreement between quasi-thermal and thermal spectrum 
in the astrophysically relevant energy range
from 6 to 10\,MeV is already obtained
by the superposition of six brems\-strahlung spectra.
The brems\-strahlung spectra have been calculated from GEANT simulations
\cite{GEANT}, which compare well with the observed
photon flux of the $^{11}$B($\gamma$,$\gamma'$) reaction
(see below). However, close to the endpoint energy the simulated spectra 
had to be slightly reduced. 
Details of the calculated
brems\-strahlung spectra will be given elsewhere \cite{Vogt00}.

For an application of the quasi-thermal brems\-strahlung spectrum
we have chosen to measure the ($\gamma$,n) cross sections of several
platinum isotopes using the activation technique.
The high sensitivity of this technique allows for a concurrent measurement
of the ($\gamma$,n) cross sections for
$^{190}$Pt (natural abundance = 0.014\%), 
$^{192}$Pt (0.782\%), and
$^{198}$Pt (7.163\%) due to similar
half-lives of the residual nuclei
$^{189}$Pt,
$^{191}$Pt,
and
$^{197}$Pt
[e.g.\ $T_{1/2}(^{191}{\rm{Pt}}) = 2.862~{\rm{d}}$].

\begin{figure}
\epsfig{figure=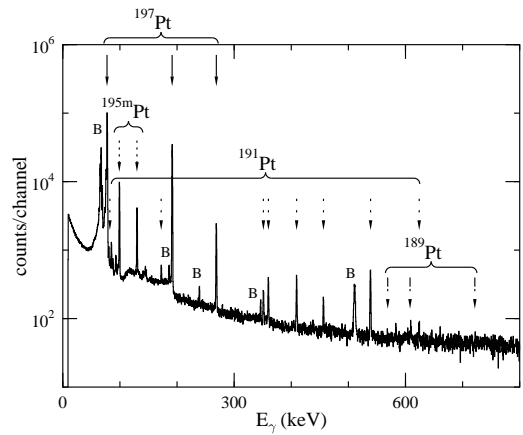,bbllx=100,bblly=85,bburx=380,bbury=305,
width=7.5cm}
%width=15.0cm}
\caption{
  \label{fig:spec_act} 
  Photon spectrum of an activated platinum disk for an endpoint
  energy of $E_{0,{\rm{max}}} = 9900$\,keV.
  The main peaks from the decay of
  $^{189}$Pt,
  $^{191}$Pt, and
  $^{197}$Pt
  and from the isomer decay of $^{195{\rm{m}}}$Pt are identified.
  Other peaks arise from room background and from X-rays (labelled B).
  The decay lines of $^{189}$Pt from the
  $^{190}$Pt($\gamma$,n)$^{189}$Pt reaction are close to the
  sensitivity limit of this experiment due to the low 
  natural abundance of $^{190}$Pt (0.014\%).
}
\end{figure}

The experiment was performed at the real photon facility of the
super\-con\-ducting Darm\-stadt
linear electron accelerator 
S--DALINAC \cite{Ric96,Mohr99,Zil00}. 
The electron beam with a typical beam current up to 40\,$\mu$A was
stopped in a massive copper disk leading to a photon flux
of about $10^5/$(keV\,cm$^{2}$\,s) at the irradiation position.
For the irradiation platinum disks (20\,mm diameter, 0.125\,mm thickness)
of natural isotopic composition 
were mounted at the target position of the ($\gamma$,$\gamma'$) setup
\cite{Mohr99}.
The disks were sandwiched between two boron layers with 
masses of about 650\,mg each. For normalization of the photon flux,
spectra of resonantly scattered photons from nuclear levels of $^{11}$B
were obtained during the activation with two high purity germanium
(HPGe) detectors (100\% relative efficiency)
placed at $90^\circ$ and $130^\circ$ relative to the incoming
photon beam (for details, see also \cite{Hart00}).
The platinum disks were irradiated for about one day, and then
they were mounted in front of a third HPGe
detector (30\% relative efficiency), where the $\gamma$-activity was observed
for one day.
A typical spectrum is shown in 
Fig.~\ref{fig:spec_act}:
$\gamma$-ray lines from the decay of the
platinum isotopes
$^{189}$Pt,
$^{191}$Pt, and
$^{197}$Pt
can clearly be identified.
Additionally, two lines from the decay of $^{195{\rm{m}}}$Pt 
were detected; this isomer is mainly populated by the
($\gamma$,$\gamma'$) reaction.
In another experiment \cite{Mohr00},
the decay curves of several lines were measured and found
to be in excellent 
agreement with the recommended half-lives \cite{NDS}.

Irradiations were performed using seven endpoint energies 
from $E_{0,{\rm{min}}} = 7200$\,keV to
$E_{0,{\rm{max}}} = 9900$\,keV in steps of 450\,keV.
In the conventional analysis of the data
the shape of the ($\gamma$,n) cross section was assumed to exhibit
a typical threshold behavior:
\begin{equation} 
\sigma = \sigma_0 \cdot \sqrt{(E-E_{\rm{thr}})/E_{\rm{thr}}} \quad \quad.
\label{eq:sqrt}
\end{equation}
The constant $\sigma_0$ was derived from the yields measured at
different endpoint energies which leads then to the
astrophysical decay rate $\lambda$ using Eq.~(\ref{eq:gamow}).
The results are summarized in Table \ref{tab:summ}.
The disadvantage of this analysis, i.e.\ the
assumed shape of the ($\gamma$,n) cross section,
can be avoided if one uses the quasi-thermal spectrum 
$\Phi_{\rm{brems}}^{\rm{qt}}(T)$
(Fig.~\ref{fig:super})
for a direct determination of the decay rate. 
In this case the integral in Eq.~(\ref{eq:gamow})
is measured directly because the experimental yield
per target nucleus $Y_i$ in the $i$-th irradiation is given by
\begin{equation} 
Y_i = \int \Phi_{\rm{brems}}(E_{0,i}) \, \sigma_{(\gamma,x)}(E) \, dE
\quad \quad.
\label{eq:yield}
\end{equation}
A comparison of Eq.~(\ref{eq:yield}) with 
Eqs.~(\ref{eq:gamow}) and (\ref{eq:approx})
relates the decay rate to the experimental
yields $Y_i$ by
\begin{equation}
\lambda(T) = \sum_i a_i(T) \cdot Y_i \quad \quad.
\label{eq:result}
\end{equation}
The average deviation
between the Planck distribution 
and the 
quasi-thermal distribution
is about 10\% in the relevant energy region around
$E_{\rm{eff}} = E_{\rm{thr}} + \frac{1}{2} kT$.
The results of the analysis with the quasi-thermal spectrum 
are also presented in Table \ref{tab:summ}
together with results from statistical model calculations using the
code NON-SMOKER \cite{Rau2000}.

For the future study of ($\gamma$,n) reactions
on $p$-nuclei with their low abundances
the sensitivity of the experiments can be improved with
enriched samples.
These ($\gamma$,n) data might restrict the relevant parameters
of $\gamma$ process models
leading eventually to
definite conclusions
for the astrophysical site of the $\gamma$ process.

Finally, under stellar
conditions the target nucleus can be excited in the thermal photon
bath reducing the effective neutron threshold by the respective
excitation energy. This two-step process must be taken into account in model
calculations \cite{Rau2000},
because in our experiment the target nucleus is always in
the ground state. 

\acknowledgements 
We thank the S--DALINAC group around H.-D.~Gr\"af
for the reliable beam during the photoactivation and
A.~Richter and U.~Kneissl for valuable discussions.
This work was supported by the Deutsche Forschungsgemeinschaft
(contracts Zi\,510/2-1 and Ri\,242/12-2). T.R.\ is supported by a PROFIL
fellowship from the Swiss National Science Foundation (grant
2124-055832.98) and by the NSF (grant NSF-AST-97-31569).

\begin{table}
\caption{\label{tab:summ} 
Summary of the ($\gamma$,n) results for the platinum isotopes
$^{190}$Pt, $^{192}$Pt, and $^{198}$Pt. From the conventional analysis 
of the data the cross section parameter $\sigma_0$ and the decay rate
$\lambda_{\rm{conv}}$ are derived. From the quasi-thermal
brems\-strahlung spectrum the decay rate 
$\lambda_{\rm{qt}}$ can be determined directly.
The experimental results are compared to a statistical model
calculation 
($\lambda_{\rm{theo}}$).
All decay rates are given for a temperature $T_9 = 2.5$.
The quoted uncertainties are the quadratic sum of statistical
and systematic (main sources:\ photon flux calibration and efficiency of the
HPGe detectors) errors \protect\cite{Vogt00}.
}
\begin{center}
\begin{tabular}{ccr@{$\pm$}lr@{$\pm$}lr@{$\pm$}lc}
target & $E_{\rm{thr}}$\,(keV)
& \multicolumn{2}{c}{$\sigma_0$\,(mb)}
& \multicolumn{2}{c}{$\lambda_{\rm{conv}}$ (s$^{-1}$)}
& \multicolumn{2}{c}{$\lambda_{\rm{qt}}$\,(s$^{-1}$)}
& $\lambda_{\rm{theo}}$ (s$^{-1}$) \\
\hline
$^{190}$Pt  & 8911 & 350 & 150  &  0.38  & 0.16 
                                   & \multicolumn{2}{c}{--}\tablenotemark[1] 
                                                      & 0.18  \\
$^{192}$Pt  & 8676 & 120 &  25  &  0.37  & 0.08   &  0.37 & 0.07   & 0.58 \\
$^{198}$Pt  & 7556 & 155 &  25\tablenotemark[2]
			  &  72    & 12\tablenotemark[3]
			     &  62   & 9\tablenotemark[3]      
				& 50\tablenotemark[3]\\
\end{tabular}
\tablenotetext[1]{
The $^{190}$Pt($\gamma$,n)$^{189}$Pt reaction could
be measured only at the highest endpoint energy $E_0 = 9900$\,keV
because of the low natural abundance of $^{190}$Pt (0.014\%).}
\tablenotetext[2]{
Our new experimental result for $\sigma_0$ of the reaction
$^{198}$Pt($\gamma$,n)$^{197}$Pt 
is in good agreement with the data from \protect\cite{Gor78} 
where the neutron yield has been measured from the GDR region down
to about 500\,keV above the ($\gamma$,n) threshold.}
\tablenotetext[3]{
The high rate of $^{198}$Pt is a consequence
of the lower threshold energy.
}
\end{center}
\end{table}

\end{multicols}

\end{document}